\documentclass[]{aa}

\usepackage{color}
\usepackage{graphicx}
\usepackage{txfonts}
\usepackage{lmodern}
\usepackage{natbib,twoopt}
\usepackage{hyperref}
\usepackage{lscape}
\usepackage{breakurl}

\hypersetup{colorlinks=true,linkcolor=blue,citecolor=blue,urlcolor=blue,breaklinks=False}

\definecolor{red}{rgb}{1,0,0}
\definecolor{blue}{rgb}{0,0,1}
\definecolor{mygreen}{RGB}{0,128,0}

\graphicspath{{pic/}}

\bibpunct{(}{)}{;}{a}{}{,} 
\makeatletter
\newcommandtwoopt{\citeads}[3][][]{\href{http://adsabs.harvard.edu/abs/#3}%
{\def\hyper@linkstart##1##2{}%
\let\hyper@linkend\@empty\citealp[#1][#2]{#3}}}
\newcommandtwoopt{\citepads}[3][][]{\href{http://adsabs.harvard.edu/abs/#3}%
{\def\hyper@linkstart##1##2{}%
\let\hyper@linkend\@empty\citep[#1][#2]{#3}}}
\newcommandtwoopt{\citetads}[3][][]{\href{http://adsabs.harvard.edu/abs/#3}%
{\def\hyper@linkstart##1##2{}%
\let\hyper@linkend\@empty\citet[#1][#2]{#3}}}
\newcommandtwoopt{\citeyearads}[3][][]%
{\href{http://adsabs.harvard.edu/abs/#3}
{\def\hyper@linkstart##1##2{}%
\let\hyper@linkend\@empty\citeyear[#1][#2]{#3}}}
\makeatother

\begin{document}

\title{The close circumstellar environment of Betelgeuse\thanks{Based on observations made with ESO telescopes at Paranal Observatory, under ESO programs 288.D-5035(A), 090.D-0548(A), 092.D-0366(A), 092.D-0366(B) and 094.D-0869 (A)}}
\subtitle{IV. VLTI/PIONIER interferometric monitoring of the photosphere}
\titlerunning{VLTI/PIONIER monitoring of Betelgeuse}
%

\author{
M.~Montarg\`es\inst{1,2}
\and
P.~Kervella\inst{2,3}
\and
G.~Perrin\inst{2}
\and
A.~Chiavassa\inst{4}
\and
J.-B.~Le~Bouquin\inst{5}
\and
M.~Auri\`ere\inst{6,7}
\and
A.~L\'opez~Ariste\inst{8,9}
\and
P.~Mathias\inst{6,7}
\and
S.~T.~Ridgway\inst{10}
\and
S.~Lacour\inst{2}
\and
X.~Haubois\inst{11}
\and
J.-P.~Berger\inst{11}
}
%
\institute{
Institut de Radioastronomie Millim\'etrique, 300 rue de la Piscine, 38406, Saint Martin d'H\`eres, France
\and
LESIA, Observatoire de Paris, PSL Research University, CNRS UMR 8109, Sorbonne Universit\'es, UPMC, Universit\'e Paris Diderot, Sorbonne Paris Cit\'e,
5 place Jules Janssen, F-92195 Meudon, France
\and
Unidad Mixta Internacional Franco-Chilena de Astronom\'{i}a (UMI 3386), CNRS/INSU, France
\& Departamento de Astronom\'{i}a, Universidad de Chile, Camino El Observatorio 1515, Las Condes, Santiago, Chile.
\and
Laboratoire Lagrange, Universit\'e C\^ote d'Azur, Observatoire de la C\^ote d'Azur, CNRS, Bd de l'Observatoire, CS 34229, 06304 Nice Cedex 4, France
\and
UJF-Grenoble 1/CNRS-INSU, Institut de Plan\'etologie et d'Astrophysique de Grenoble (IPAG), UMR 5274, Grenoble, France
\and
Universit\'e de Toulouse, UPS-OMP, IRAP, Tarbes, France
\and
CNRS, IRAP, 57 Avenue d'Azereix, BP 826, F-65008 Tarbes Cedex, France
\and
Universit\'e de Toulouse, UPS-OMP, Institut de Recherche en Astrophysique et Plan\'etologie, Toulouse, France
\and
CNRS, UMR5277, Institut de Recherche en Astrophysique et Plan\'etologie, 14 Avenue Edouard Belin, 31400 Toulouse, France
\and
National Optical Astronomy Observatories, 950 North Cherry Avenue, Tucson, AZ 85719, USA
\and
European Southern Observatory, Alonso de Cordova 3107, Casilla 19001, Vitacura, Santiago 19, Chile
}

\date{Received 2015 July 21 / Accepted 2016 February 07}

\abstract
        {The mass-loss mechanism of cool massive evolved stars is poorly understood. The proximity of Betelgeuse makes it an appealing target to study its atmosphere, map the shape of its envelope, and follow the structure of its wind from the photosphere out to the interstellar medium.}
        {A link is suspected between the powerful convective motions in Betelgeuse and its mass loss. We aim to constrain the spatial structure and temporal evolution of the convective pattern on the photosphere and to search for evidence of this link.}
        {We report new interferometric observations in the infrared $H$ band using the VLTI/PIONIER instrument. We monitored the photosphere of Betelgeuse between 2012 January and 2014 November to look for evolutions that may trigger the outflow.
        }
        {Our interferometric observations at low spatial frequencies are compatible with the presence of a hot spot on the photosphere that has a characteristic width of one stellar radius. It appears to be superposed on the smaller scale convective pattern. In the higher spatial frequency domain, we observe a significant difference between the observations and the predictions of 3D hydrodynamical simulations.}
        {We bring new evidence for the presence of a convective pattern in the photosphere of red supergiants. The inferred hot spot is probably the top of a giant convection cell although an asymmetric extension of the star cannot be excluded by these interferometric observations alone. The properties of the observed surface features show a stronger contrast and inhomogeneity as predicted by 3D radiative hydrodynamical simulations. We propose that the large observed feature is modifying the signature of the convective pattern at the surface of the star in a way that simulations cannot reproduce.}

\keywords{Stars: individual: Betelgeuse; Stars: imaging; Stars: supergiants; Stars: mass-loss; Infrared: Stars, Techniques: interferometric}

\maketitle

\section{Introduction \label{Sect_Intro}}

Evolved stars are important contributors to the enrichment of the interstellar medium in heavy elements and dust, and more generally to the chemical evolution of the Universe. In starburst galaxies or galaxies at large look-back times,  red supergiants (RSG) are usually expected to be the main source of dust. 

Thanks to its proximity \citepads[$\approx 197$\,pc, ][]{2008AJ....135.1430H}, \object{Betelgeuse} ($\alpha$\,Ori, \object{HD 39801}, \object{HR 2061}) has one of the largest apparent photospheric diameters (42 milliarcseconds (mas) in the infrared $K$ band, \citeads{2014A&A...572A..17M}), converting to an extremely large linear radius of the order of 1000\,R$_\odot$. It is a M2Iab prototypical RSG, with a mass estimated between $\mathrm{11.6_{-3.9}^{+5.0}}$\,M$_\odot$ \citepads{2011ASPC..451..117N} and 15-20\,M$_\odot$ \citepads{2008nuco.confE.242D}. These favorable observational properties make it one of the best studied RSG, although its extreme brightness ($m_K < -4$) paradoxically makes it  a difficult target for large telescopes, as it often causes detector saturation problems.

Mass loss of these stars remains poorly understood. In particular, the process triggering the outflow is unknown as they do not experience flares or pulsations that are large enough. RSG are semi-regular variable. Many of them have one or two photometric and/or radial velocity periods (\citeads{2006MNRAS.372.1721K}, and \citeads{2012ApJ...754...35Y}). The short periods are of a few hundred days and of a few thousand days for the long secondary period (LSP). In the case of Betelgeuse, these periods are respectively of $\sim 400$ and $\sim 2100$~days \citepads{2010ApJ...725.1170S}. The short period is assumed to correspond to radial amplitude pulsations, whereas the nature of the LSP remains unexplained. \citetads{2010ApJ...725.1170S} suggests that it could be linked to giant convection cell turnover. \citetads{2015A&A...575A..50A} show that the pulsation models cannot account for the extension of the molecular layers of RSG and, thus, that pulsations cannot be responsible for the mass loss. In particular, the particle velocity does not meet the escape velocity ($\sim$~65-90 km.s$^{-1}$ in the case of a non-rotating Betelgeuse). \citetads{2007A&A...469..671J} obtained spectroscopic observations of a sample of RSGs and concluded that radiative pressure on molecular lines, together with convection that decreases the effective gravity, could initiate the mass loss. However, the presence of a weak magnetic field on Betelgeuse \citepads{2010A&A...516L...2A} may play a role in the mass-loss mechanism of that kind of star, e.g. through the dissipation of Alfv\'en waves: \citetads{2000ApJ...528..965A} developed a 2.5 MHD (magnetohydrodynamics) code that reproduces the terminal wind velocity and the measured mass-loss rate of Betelgeuse. This indicates that Alv\' en waves can actually drive the outflow of massive cool evolved stars.

\citeads{2009A&A...504..115K}, hereafter~Paper~I, obtained adaptive optics (AO) images of Betelgeuse with the NACO (Nasmyth Adaptive Optics System coupled with the CONICA camera) instrument at the VLT (Very Large Telescope) in the spectral range $1.04-2.17\,\mu$m. They revealed the presence of  several ``plumes" that extend up to a radius of six times the photosphere, most probably of a gaseous nature and containing the CN molecule. In the $K_s$ band ($\lambda \approx 2.2\,\mu$m), \citeads{2011ApJ...743..178M} also measured a significant resolved emission around Betelgeuse (astrophysical null depth of $8.1 \times 10^{-2} \pm 4 \times 10^{-3}$) using the Palomar Fiber Nuller instrument. Diffraction-limited VLT/VISIR (VLT Imager and Spectrometer for mid Infrared) observations (\citeads{2011A&A...531A.117K}, hereafter Paper II), in the thermal infrared domain ($7.8-19.5\,\mu$m) uncovered a large, clumpy circumstellar nebula, most likely made of O-rich dust that formed from the material expelled by Betelgeuse. Finally, using the recently commissioned SPHERE (Spectro-Polarimetric High-contrast Exoplanet REsearch) instrument at the VLT in the visible, \citetads{2016A&A...585A..28K}, hereafter~Paper~III, managed to resolve the apparent disk of Betelgeuse. In narrow band filters, the intensity images revealed a departure from spherical symmetry. The polarized signal indicates the possible presence of dust very close to the star (less than one stellar radius) in the northeastern quadrant. These observations suggest that the mass loss of the star is episodic and non-spherical.

A convective pattern was identified on Betelgeuse \citepads{2010A&A...515A..12C}, as well as bright spots on the photosphere \citepads{2009A&A...508..923H}. Using the Berkley Infrared Spatial Interferometer (ISI) between 2006 and 2010, \citetads{2011ApJ...740...24R} monitored the circumstellar environment of this star at 11.5~$\mu$m. For some epoch they observed, they detected bright point sources on the layer they modeled at an optical depth of one.
However, \citetads{2015A&A...575A..50A} showed that 3D radiative hydrodynamical simulations cannot account for the molecular extension of the atmosphere of RSG. Only a few interferometric snapshots of RSG have been obtained up until now, sampling a very limited spatial frequency range and sometimes only one azimuthal direction. These observations were interpreted using various kinds of models \citepads[see~e.g. ][]{1997MNRAS.290L..11B, 2000MNRAS.315..635Y, 2004A&A...418..675P, 2006sf2a.conf..471H}. Image reconstructions were also performed using observations with a sufficiently dense ($u,v$) coverage \citepads{1985ApJ...295L..21R,2009A&A...508..923H}. Taking advantage of the four-telescope beam combiner PIONIER (Precision Integrated-Optics Near-infrared Imaging ExpeRiment) at the Very Large Telescope Interferometer (VLTI), we obtained multi-directional ($u,v$) coverage with four epochs of observations (Sect.~\ref{Sect_Observations}). This homogeneous dataset enabled us to monitor the photosphere of a RSG for the first time. We present a classical model-based analysis of these new data (Sect.~\ref{Sect_LowSF}) and a comparison of the data with numerical radiative-hydrodynamic (RHD) simulations (Sect.~\ref{Sect_RHD}).

\section{Observations \label{Sect_Observations}}

The European Southern Observatory's VLTI \citepads{2010SPIE.7734E...3H} has been operating  on top of the Cerro Paranal, in Northern Chile, since March 2001. We observed Betelgeuse and its associated calibrators on the nights of 2012 January 31, 2013 February 09, 2014 January 11, 2014 February 01, and 2014 November 21 (local time) using the VLTI equipped with the PIONIER instrument \citepads{2011A&A...535A..67L}. The log of the observations is presented in Table \ref{observations_log}. PIONIER is an innovative interferometric beam-combiner that relies on an integrated optics component \citepads{2009A&A...498..601B} for the recombination of the light beams. It saw first light at the VLTI in 2010, and was installed as a visitor instrument until Spring 2015 (starting in P96, PIONIER is now offered as a facility instrument). We used the high spectral resolution of the dispersion unit, that produces seven spectral channels across the $H$ band (with a spectral resolution of approximately 40) for the 2012 and 2013 observations. For the 2014 observations, we used the same spectral resolution, reading only the three central pixels of the detector.

\begin{table}[ht!]
        \caption{Log of the PIONIER observations of Betelgeuse and its calibrators}
        \label{observations_log}
        \centering
        \begin{tabular}{llll}
                \hline \hline
                \noalign{\smallskip}
                UT & & Star & Configuration\\
                \hline
                \noalign{\smallskip}
                2012 Feb-01 & 01:07 & Sirius A & A1-B2-C1-D0\\
                 & 01:17 & Sirius A & A1-B2-C1-D0\\
                 & 01:53 & Betelgeuse & A1-B2-C1-D0\\
                 & 02:10 & Sirius A & A1-B2-C1-D0\\
                 & 02:34 & Betelgeuse & A1-B2-C1-D0\\
                 & 03:08 & Sirius A & A1-B2-C1-D0\\
                 & 03:20 & Sirius A & A1-B2-C1-D0\\
                 & 03:36 & Betelgeuse & A1-B2-C1-D0\\
                 & 03:50 & Sirius A & A1-B2-C1-D0\\
                 & 04:05 & Betelgeuse & A1-B2-C1-D0\\
                2013 Feb-10 & 00:25 & b Ori & A1-B2-C1-D0\\ 
                 & 00:51 & Betelgeuse & A1-B2-C1-D0\\
                 & 01:09 & LTT 11688 & A1-B2-C1-D0\\ 
                 & 01:27 & Betelgeuse & A1-B2-C1-D0\\ 
                 & 01:45 & b Ori & A1-B2-C1-D0\\
                 & 02:05 & Betelgeuse & A1-B2-C1-D0\\
                 & 02:28 & LTT 11688 & A1-B2-C1-D0\\   
                 & 02:47 & Betelgeuse & A1-B2-C1-D0\\
                 & 03:08 & b Ori & A1-B2-C1-D0\\
                 & 03:19 & Betelgeuse & A1-B2-C1-D0\\
                 & 03:37 & 56 Ori & A1-B2-C1-D0\\
                2014 Jan-12 & 01:55 & LTT 11688 & A1-B2-C1-D0\\ 
                 & 02:18 & Betelgeuse & A1-B2-C1-D0\\
                 & 02:32 & b Ori & A1-B2-C1-D0\\
                 & 02:46 & Betelgeuse & A1-B2-C1-D0\\
                 & 02:57 & 56 Ori & A1-B2-C1-D0\\
                 & 03:09 & Betelgeuse & A1-B2-C1-D0\\
                 & 03:20 & LTT 11688 & A1-B2-C1-D0\\
                2014 Feb-02 & 03:04 & b Ori & A1-G1-K0-J3\\
                 & 03:20 & Betelgeuse & A1-G1-K0-J3\\
                 & 03:34 & LTT 11688 & A1-G1-K0-J3\\
                2014 Nov-22 & 04:32 & LTT 11688 & A1-B2-C1-D0\\
                 & 05:00 & Betelgeuse & A1-B2-C1-D0\\
                 & 05:57 & Betelgeuse & A1-B2-C1-D0\\
                 & 06:19 & LTT 11688 & A1-B2-C1-D0\\
                 & 06:32 & Betelgeuse & A1-B2-C1-D0\\
                 & 06:46 & 56 Ori & A1-B2-C1-D0\\
                 & 06:59 & Betelgeuse & A1-B2-C1-D0\\
                 & 07:15 & LTT 11688 & A1-B2-C1-D0\\
                 & 07:44 & Betelgeuse & A1-B2-C1-D0\\
                 & 07:57 & b Ori & A1-B2-C1-D0\\
                 & 08:08 & Betelgeuse & A1-B2-C1-D0\\
                 & 08:22 & LTT 11688 & A1-B2-C1-D0\\
                 & 08:37 & Betelgeuse & A1-B2-C1-D0\\
                 & 08:54 & 56 Ori & A1-B2-C1-D0\\
                \hline
        \end{tabular}
\end{table}

We used the four auxiliary telescopes (ATs, 1.8\,m aperture) installed on the A1-B2-C1-D0 and A1-G1-K0-J3 baseline quadruplets. These AT configurations were the most compact and the more extended offered by ESO at that time, with baseline lengths (measured on the ground) between 11.3 and 35.8\,m, for the first, and between 80.0 and 153.0\,m, for the second. The resulting $(u,v)$ coverage is presented in Fig.~\ref{Fig:v2_cp_hotspot}. The raw data were processed using the publicly available PIONIER pipeline \citepads{2011A&A...535A..67L} to produce six squared visibilities ($V^2$) and four independent closure phase (CP) values per observation, for each spectral channel. Uncertainties on uncalibrated observables are computed from the statistical dispersion over the 100 scans inside each $\sim 30$~s exposure. The uncertainty on the transfer function is then added quadratically.
        
        
To avoid detector saturation caused by the high brightness of Betelgeuse, we lowered the incoming flux. In 2012, diaphragms were used in the beams. During this run, we observed the first zero of the visibility function at two different spatial frequencies depending on the observed ($u, v$) direction (i.e. depending on the baseline). To exclude an effect of the diaphragms on the coherence of the beam, we inserted a neutral density in front of the detector in 2013. In 2014, following an update of the instrument's software, we used the ability to read only the three central pixels of the seven available with the large dispersion. This allows  the length of the reading time to be reduced. 
        
The choice of the calibrator star to monitor the interferometric transfer function of the instrument is complicated by the extreme brightness of Betelgeuse. Following \citeads{2009A&A...503..183O, 2011A&A...529A.163O}, we chose to rely on \object{Sirius~A} on 2012 January, as its brightness is comparable to that of Betelgeuse in the $H$ band ($m_H=-1.4$, vs. $m_H=-4.0$), resulting in a similar signal-to-noise ratio with the same PIONIER setup. Owing to its much higher effective temperature, Sirius~A has a comparatively small angular diameter of $\theta_\mathrm{LD}=6.039 \pm 0.019$\,mas \citepads{2003A&A...408..681K}, which results in a high fringe visibility on the selected baselines, and a good quality transfer function calibration. Moreover, Sirius~A is not expected to have significant surface features, and is therefore also a good choice in terms of closure phase calibration. However, in 2012, the instrument team learned that the transfer function of PIONIER is sky position dependent. As a result, considering the unusual first lobe shape of our data (Fig.~\ref{Fig:v2_cp_hotspot}) and the distance between Betelgeuse and Sirius A on the sky ($\sim 27^\circ$), we decided to choose closer calibrators for the 2013 February, 2014 January, February, and November runs. Stars at 7$^\circ$ or less from Betelgeuse were used: \object{56 Ori}, \object{b Ori,} and \object{LTT 11688} (Table \ref{Tab:calibrators_data}). 

\begin{table}[ht!]
        \caption{List of calibrators. Angular distances were measured using Aladin. }
        \label{Tab:calibrators_data}
        \centering
        \begin{tabular}{llll}
                \hline \hline
                \noalign{\smallskip}
                Name & Angular dist. & Diameter & Ref.\\
                & from $\alpha$ Ori ($^\circ$) & (mas) & \\
                \hline
                \noalign{\smallskip}
                Sirius A & 27 & $6.039 \pm 0.019$ & 1 \\
                56 Ori & 6 & $2.38 \pm 0.044$ & 2 \\
                b Ori & 7 & $1.78 \pm 0.13$ & 3 \\
                LTT 11688 & 5 & $2.08 \pm 0.15$ & 3 \\
                \hline
        \end{tabular}
        \tablebib{(1)~\citetads{2003A&A...408..681K}; (2)~\citetads{2002A&A...393..183B}; (3)~\citetads{2006A&A...456..789B,2011A&A...535A..53B}}
\end{table}

The data of the 2014 February 02 on the extended configuration of the ATs of the VLTI probed a high spatial frequency domain (from 151 to 330 arcsec$^{-1}$, meaning from the 6th to the 15th lobe of visibility). The weather conditions were not good enough (dust in the atmosphere, seeing $> 1"$) to obtain reliable data. Therefore, we did not consider these data any further. 

Considering the spectral resolution of the PIONIER instrument, we  use all the spectral elements together in the present work, as a "pseudo-continuum".

\section{Low spatial frequencies: analytical model fitting \label{Sect_LowSF}}

\subsection{Doubling of the first lobe \label{SubSect_first_lobe}}

Our data present an unusual azimuthal dependence of the spatial frequency at which the first zero of the visibility function occurs (Fig. \ref{Fig:v2_cp_hotspot}, left column). During our four epochs of observations, we changed the settings of the instrument. In particular, we used three different features to avoid detector saturation (diaphragms, neutral density, and acquisition of three spectral channels, instead of seven). The separated first zeros remain. Thus it supports the theory that this is not an instrumental artifact but a real characteristic of the star.

This kind of feature indicates that the star does not appear as a disk for the interferometer, as its angular diameter depends on the observed direction on the plane of the sky. Therefore, usual uniform disk (UD) or limb-darkened disk (LDD) models cannot reproduce this data. An elliptical model would seem appropriate. However, the strong closure phase signal (Fig. \ref{Fig:v2_cp_hotspot}, right column) excludes this kind of  model without asymmetries, since its closure phases are at 0$^\circ$ or 180$^\circ$.

\subsection{Limb-darkened disk and Gaussian hot spot \label{SubSect_HotSpot}}

\subsubsection{Model definition}

As indicated in Sect. \ref{SubSect_first_lobe}, the first lobe of the visibility function does not have the same extension that depends on the position angle (PA). \citetads{2009A&A...506.1351C,2010A&A...515A..12C} have already shown that huge and bright convection cells could create some dispersion of the spatial frequency at which the first zero of the visibility function occurs when considered as a function of the PA (as high as 5\% on both the first zero position and on the determination of the stellar radii). As these structures were previously observed on Betelgeuse at smaller scales (\citeads{1990MNRAS.245P...7B}; \citeads{1992MNRAS.257..369W,1997MNRAS.291..819W}; \citeads{1997MNRAS.285..529T}; \citeads{2000MNRAS.315..635Y}; \citeads{2009A&A...508..923H}), we decided to use a model that combines a stellar disk and a single bright spot.

Recent interferometric observations of Betelgeuse (\citeads{2009A&A...503..183O,2011A&A...529A.163O}, and \citeads{2014A&A...572A..17M}) were analyzed using a power-law LDD model \citepads{1997A&A...327..199H}. Therefore, to compare our results, we used the same model for the photosphere and a single 2D symmetrical Gaussian for the bright hotspot. The center of this hotspot is located at the coordinates ($x_\mathrm{center}$, $y_\mathrm{center}$) respectively, in right ascension and declination, relative to the center of the stellar disk. The size of this Gaussian spot is given by its full width at half maximum (FWHM). The relative weight of both structures are $w_\mathrm{LDD}$ and $w_\mathrm{spot}$ , which are defined so that

\begin{equation}
        w_\mathrm{LDD} + w_\mathrm{spot} = 1.
\end{equation}

Given the linearity of the Fourier transform, the model visibility is

\begin{equation}
        V_\mathrm{model} = w_\mathrm{LDD}V_\mathrm{LDD} + w_\mathrm{spot}V_\mathrm{spot}.
\end{equation}

\noindent $V_\mathrm{LDD}$ is given by

\begin{equation}
        V_\mathrm{LDD}(u,v) = \Gamma(\nu+1)\frac{J_\nu (x)}{(x/2)^\nu}\label{Eq:VisLDD}
,\end{equation}

\noindent with $x = \pi \theta_\mathrm{LDD} \sqrt{u^2+v^2}$, $\nu = \alpha/2 +1$, $\alpha$ being the limb-darkening parameter, $J_\nu$ is the first species Bessel function of order $\nu$, and $\Gamma$ the Euler function. The complex visibility of the hot spot is then

\begin{equation}
        \begin{array}{l}
                V_\mathrm{spot}(u, v) = \exp\left[-\frac{( 2\pi f \sigma)^2} {2} \right]\\\\
                \times \exp\left[-2i\pi(u x_\mathrm{center} + v y_\mathrm{center})\right]
        \end{array}
\end{equation}
        
\noindent with $f = \sqrt{u^2 + v^2}$ and $\sigma = $~FWHM$/(2\sqrt{2\ln(2)})$.\\

We note that, with this model, light distribution is allowed to extend beyond the stellar disk as the Gaussian spot has an infinite extension. For hotspots near the center of the disk this has little consequence, but it is not negligible when the feature is close to the limb of the star. 

\subsubsection{Model fitting}

The model has six parameters: $\theta_\mathrm{LDD}$ the limb-darkened disk diameter, $\alpha_\mathrm{LDD}$ the limb darkening power law exponent, $w_\mathrm{spot}$, $x_\mathrm{center}$, $y_\mathrm{center}$, and the FWHM. To fit the data, we need an initial guess for these parameters. This task is made complex by the strong indeterminacy of the apparent diameter that is caused by the presence of the Gaussian spot (variations of the first lobe of the visibility function).\\

Therefore, we decided to fit our model in two steps. First we explored the parameter space by using $\chi^2$ maps, and then we performed a regular fit using the least-squares method.

We fitted both the squared visibilities and the closure phases for spatial frequencies lower than 51~arcsec$^{-1}$, selecting the baselines and triplets that sample only the first and second lobes to avoid contamination by small scale structures. The $\chi^2$ maps were computed by using a grid of spot positions ($x_\mathrm{center}$, $y_\mathrm{center}$) with a 2~mas step between -25 and 25~mas, relative to the center of the star. We excluded positions with the hotspot outside the star. On each cell of the grid, we fitted $\theta_\mathrm{LDD}$, $\alpha_\mathrm{LDD}$, and $w_\mathrm{spot}$. The initial guesses for the star were the LDD parameters from \citetads{2014A&A...572A..17M}, which are the closest value we have in time in the near infrared, even though they used $K$-band observations. The $w_\mathrm{spot}$ was initially set at 0.2. The maps were computed for spot FWHM fixed at 6, 10, 15, 20, and 25~mas and, for each FWHM, we selected the cell offering the minimum $\chi^2$. For each epoch we then used the corresponding parameters as initial guesses for our fit on all  six parameters.

\subsubsection{Results}

\begin{figure*}
        \centering
        \resizebox{\hsize}{!}{\includegraphics{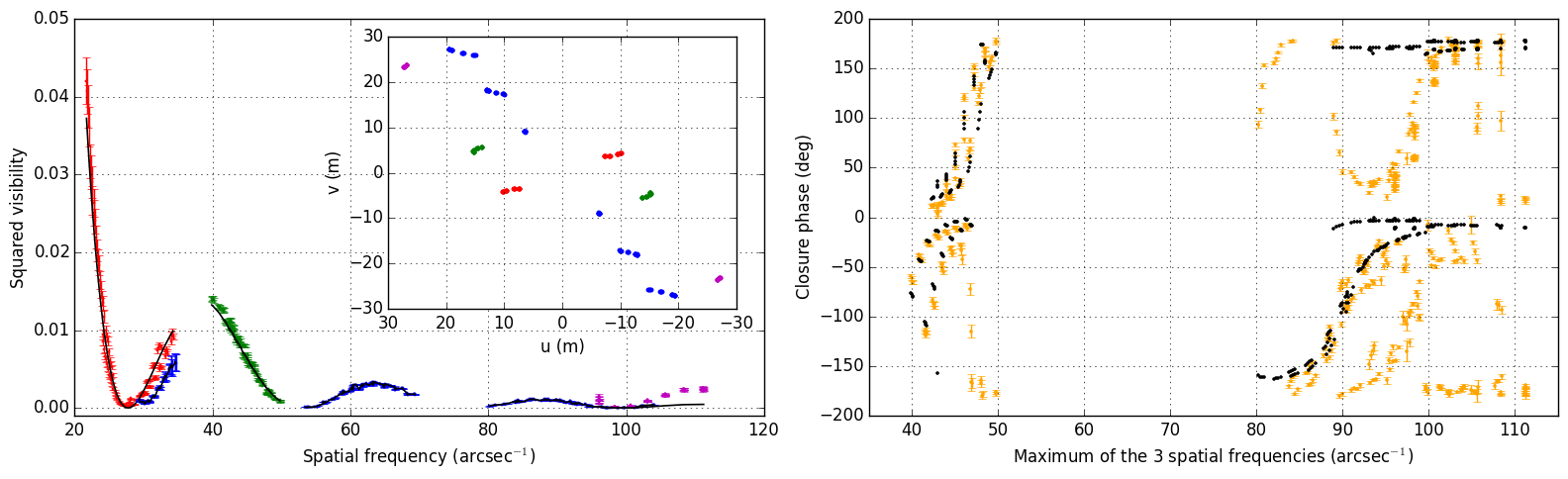}}\\
        \resizebox{\hsize}{!}{\includegraphics{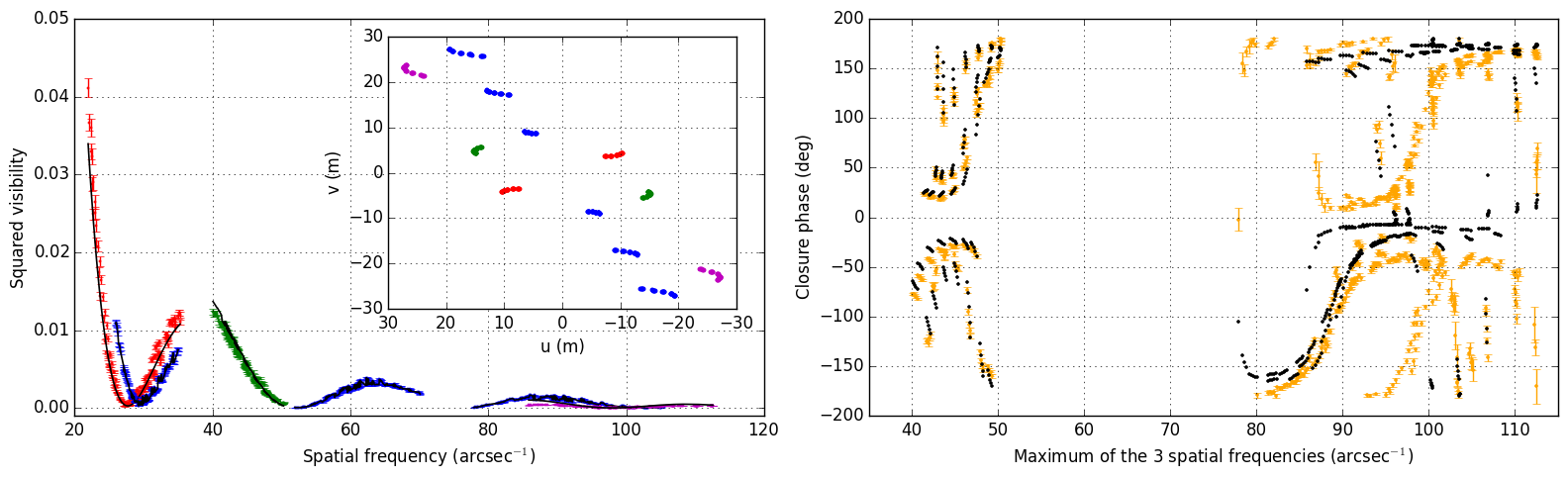}}\\
        \resizebox{\hsize}{!}{\includegraphics{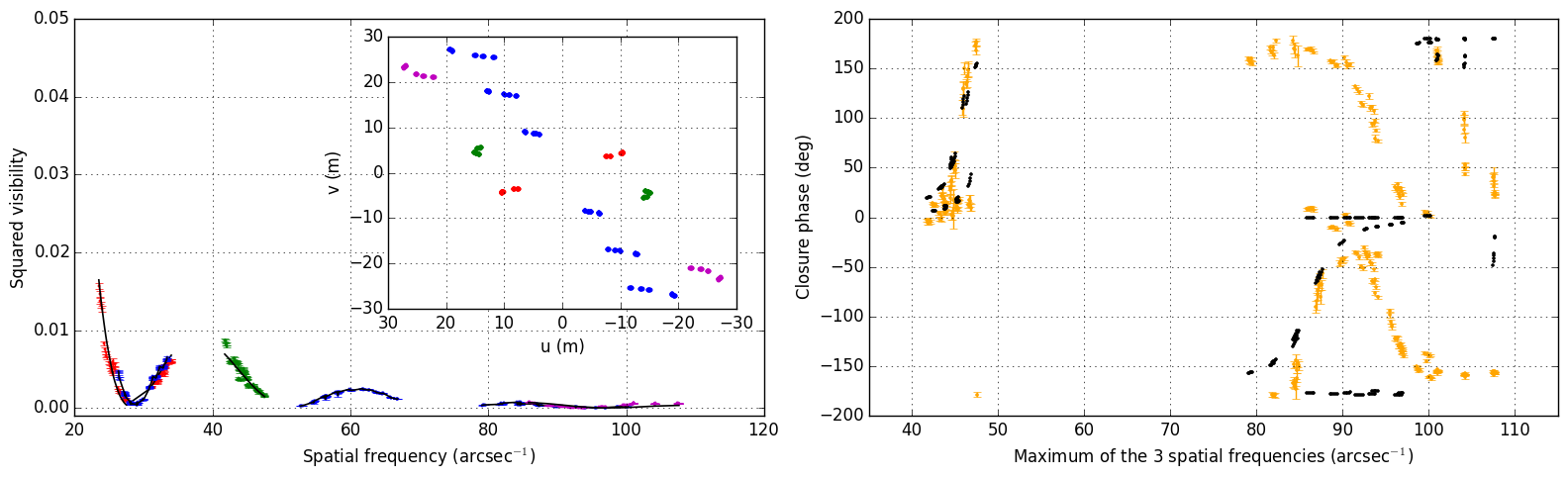}}\\
        \resizebox{\hsize}{!}{\includegraphics{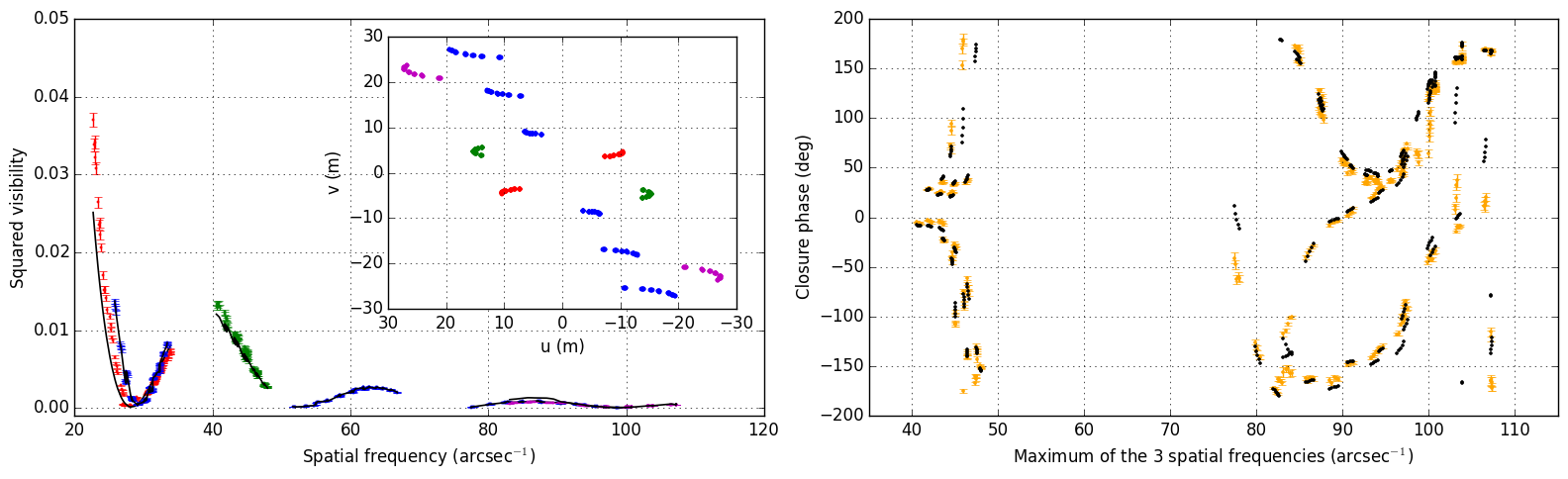}}\\
        \caption{Fit of the PIONIER data by the LDD and hotspot model. Only spatial frequencies lower than 51 arcsec$^{-1}$ were considered. \textit{Insets in the left column:} PA color-coded  ($u,v$) coverage. North is up and East is left. \textit{Left column:} squared visibilities with matching colors. \textit{Right column:} closure phases. The best-fitted model is represented in black. The four rows correpond to the 2012 January epoch (\textit{first row}), 2013 February epoch (\textit{second row}), 2014 January epoch (\textit{third row}), and 2014 November epoch (\textit{fourth row}). For this latter epoch, a two-spots model was used and the whole spatial frequency range was considered. \label{Fig:v2_cp_hotspot}}
\end{figure*}

The results of the model-fitting are presented in Table \ref{Tab:Result_fit_hotspot} and on Fig. \ref{Fig:v2_cp_hotspot}. In the following, we  use $\tilde{\chi}^2$ for the reduced $\chi^2$. The error bars were estimated using
\begin{equation}
        \chi^2(\mathrm{param} + \sigma) = 2 \chi^2_\mathrm{min}. \label{Eq:error}
\end{equation}

\begin{table*}[]
        \caption{LDD and Gaussian hotspot model fitting for the spatial frequencies lower than 51 arcsec$^{-1}$.}
        \label{Tab:Result_fit_hotspot}
        \centering
        \begin{tabular}{llll}
                \hline\hline
                \noalign{\smallskip}
                Parameter & Jan 2012 & Feb 2013 & Jan 2014 \\
                \hline
                \noalign{\smallskip}
                $\theta_\mathrm{LDD}$ (mas) & $42.94 \pm 0.50$ & $43.73 \pm 0.50$ & $44.06 \pm 0.59$ \\
                $\alpha_\mathrm{LDD}$ & $0.15 \pm 0.06$ & $0.19 \pm 0.07$ & $0.23 \pm 0.09$ \\
                $w_\mathrm{spot}$  & $0.13 \pm 0.03$ & $0.08 \pm 0.02$ & $0.22 \pm 0.08$ \\
                $x_\mathrm{center}$ (mas) & $17.07 \pm 2.22$ & $19.76 \pm 2.02$ & $16.24 \pm 5.25$ \\
                $y_\mathrm{center}$ (mas) & $-5.98 \pm 2.42$ & $-7.46 \pm 2.42$ & $2.40 \pm 5.25$ \\
                FWHM (mas) & $21.76 \pm 2.02$ & $18.42 \pm 2.42$ & $29.48 \pm 6.26$ \\
                $\tilde{\chi}^2$ & 29 & 31 & 29 \\
                $\sigma(\chi^2)/\sqrt{2\ \mathrm{dof}}$ & 46 & 33 & 30 \\
                \hline
        \end{tabular}
        \tablefoot{The last line corresponds to the standard deviation of the $\chi^2$ divided by the square root of twice the number of degree of freedom.}
\end{table*}

For the 2014 November epoch, it was not possible to converge on an unique solution: two different positions of the Gaussian hotspot gave nearly the same $\tilde{\chi}^2$. We tried to fit a two-spot model on this particular epoch by using the ambiguous derived positions as initial guesses. The fitting procedure converged toward a resolved hotspot plus a non-resolved one. Therefore, we performed this fitting process again, using the whole spatial frequency range. The results of this fit are presented in Table \ref{Tab:fit_Nov14} and in the lowest row of Fig. \ref{Fig:v2_cp_hotspot}. 

\begin{table}[ht!]
        \centering
        \caption{LDD and 2 Gaussian hotspots model-fitting for the whole spatial frequency range of the 2014 November epoch.}
        \label{Tab:fit_Nov14}
        \begin{tabular}{ll}
                \hline\hline
                \noalign{\smallskip}
                Parameters & Best-fit values \\
                \hline
                \noalign{\smallskip}
                $\theta_\mathrm{LDD}$ (mas) & $43.15 \pm 0.50$ \\
                $\alpha_\mathrm{LDD}$ & $0.14 \pm 0.09$ \\
                $w_\mathrm{spot 1}$ & $0.10 \pm 0.07$ \\
                $x_\mathrm{center 1}$ (mas) & $-21.55\pm 6.26$ \\
                $y_\mathrm{center 1}$ (mas) & $6.52 \pm 6.46$ \\
                FWHM$_\mathrm{spot 1}$ (mas) & $24.62 \pm 8.08$ \\
                $w_\mathrm{spot 2}$ & $(6.98 \pm 2.56)\times 10^{-3}$ \\
                $x_\mathrm{center 2}$ (mas) & $13.36 \pm 1.41$ \\
                $y_\mathrm{center 2}$ (mas) & $-19.07 \pm 1.41$ \\
                FWHM$_\mathrm{spot 2}$ (mas) & $4.00 \times 10^{-4} ~ ^{+2.62}_{-4.00 \times 10^{-4}}$ \\
                $\tilde{\chi}^2$ & 70 \\
                $\sigma(\chi^2)/\sqrt{2\ \mathrm{dof}}$ & 97 \\
                \hline
        \end{tabular}
        \tablefoot{The last line corresponds to the standard deviation of the $\chi^2$, divided by the square root of twice the number of degree of freedom.}
\end{table}

The best-fit LDD and hotspot model reproduce the data. The $\tilde{\chi}^2$ has its value between 29 and 31 for the first three epochs. By fitting only the spatial frequencies below 51 arcsec$^{-1}$, we manage to reproduce the visibilities up to 90 arcsec$^{-1}$, and part of the closure phases up to 110 arcsec$^{-1}$ (Fig. \ref{Fig:v2_cp_hotspot}).

Fitting the two-spot model on the other epochs gave bad results: either the two-spot solution was not significantly better ($\chi^2$ not lower, the shape of the observables not well reproduced, particularly for the first lobe of the visibility function), or $w_\mathrm{spot 2}$ converged to zero.

\begin{figure}
        \centering
        \resizebox{\hsize}{!}{\includegraphics{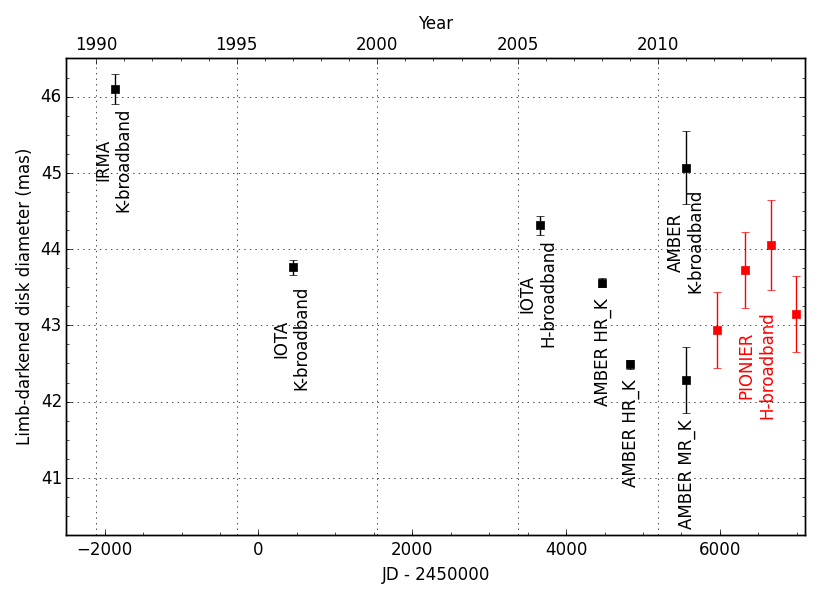}}
        \caption{Evolution of the near infrared diameter of Betelgeuse over the  past two decades. The IRMA measurement comes from \citetads{1992AJ....104.1982D}, IOTA from \citetads{2004A&A...418..675P} in K band and \citetads{2009A&A...508..923H} in H band, the two AMBER high-resolution K-band measurements from \citetads{2009A&A...503..183O,2011A&A...529A.163O}, the AMBER K-medium resolution and broadband from \citetads{2014A&A...572A..17M} and the PIONIER H-band measurements are from the present work.}
        \label{Fig:Diam_Evol}
\end{figure}

The spotty model also gives  an LDD diameter and an LD parameter which are typical for the much observed star Betelgeuse. This was not the case for the LDD disk model alone. Figure \ref{Fig:Diam_Evol} represents the archival near-infrared LDD diameter measured on Betelgeuse for the last two  decades. Our values are in the same range as previous measurements ($46.1 \pm 0.2$~mas from \citeads{1992AJ....104.1982D}, $43.76 \pm 0.12$~mas from \citeads{2004A&A...418..675P}, $44.31 \pm 0.12$~mas from \citeads{2009A&A...508..923H}, 43.6~mas from \citeads{2009A&A...506.1351C}, $43.56 \pm 0.06$~mas from \citeads{2009A&A...503..183O}, $42.49 \pm 0.06$~mas from \citeads{2011A&A...529A.163O}, and $42.28 \pm 0.43$~mas for the K continuum, and $45.07 \pm 0.48$~mas for the whole K band, both from \citeads{2014A&A...572A..17M}).

The best-fit models for the four epochs give $\tilde{\chi}^2$ values greater than one and the standard deviation of the $\chi^2$ is much greater than the square root of twice the number of degree of freedom. However, the shape of the derived squared visibilities and closure phases reproduce the
observed data well. This comes from the chosen model. Indeed, we decided to consider symmetric Gaussian hotspots. The shape of this structure is certainly more complex than the structure we adopted and, as a consequence, it must have an effect on the observables, particularly the closure phases. We could have used an elliptical gaussian to simply mimic this supposed behavior, but this would have added two variables in a very degenerated parameter space and led to larger uncertainties on the best-fitted model. Moreover the solution would certainly not have been unique, as only a few changes in the high spatial frequency domain in the Fourier space would have been sufficient to improve the fit, without significantly altering the resulting direct image. This seems characteristic of a sparse sampling of the $uv$ plane for a largely resolved object.

The main obstacle is that we cannot derive the star diameter with an independent process: the larger hotspot has a direct effect on the first lobe of the visibility function, usually used for this process.

\citetads{2009A&A...508..923H} fitted their data with two hotspots. The largest one remains two times smaller than ours. Their positions on the stellar disk are different, which is to be expected as their observations took place in 2005. Indeed, according to radiative-hydrodynamics simulations, bright and large convective cells being seen in the near infrared evolve on a timescale of years (the simulations explore several years of stellar evolution and, in the produced images, we directly see that large and bright hotspots take at least a year to evolve significantly). We note that neither these authors nor we observe a counterpart to the chromospheric bright spot detected by \citetads{1998AJ....116.2501U} and tentatively identified as the southern pole of the star. This difference was addressed by \citetads{2011IAUS..273..188D}: the formation of the infrared and ultraviolet structures are certainly very different. 

\begin{figure}
        \centering
        \resizebox{\hsize}{!}{\includegraphics{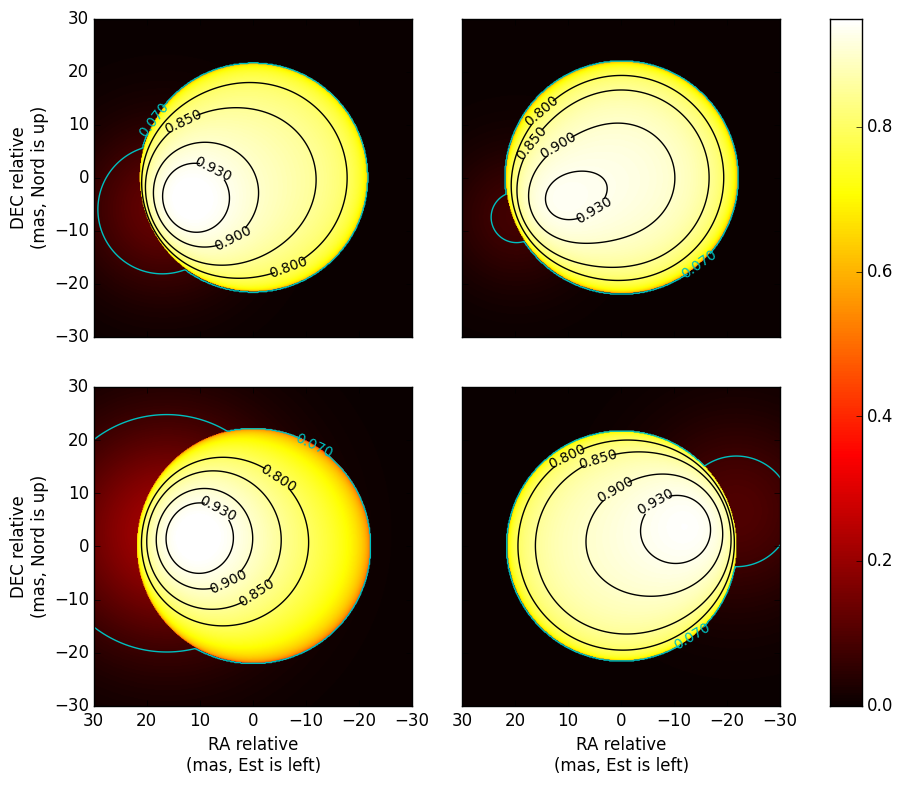}}
        \caption{Intensity maps derived from the best fitted LDD and Gaussian hotspot model. North is up and East to the left.}
        \label{Fig:Int_maps}
\end{figure}

We would like to emphasize that it is the four-telescope ($u, v$) coverage of PIONIER that allowed us to sample two directions almost orthogonal in the Fourier plane. Without this, the displacement of the spatial frequency at which the first null of the visibility function occurs would probably have been interpreted as an increase of the stellar diameter.

\section{Radiative-hydrodynamics simulations \label{Sect_RHD}}

Radiative-hydrodynamics simulations (RHD) simulations have been previously used to interpret the interferometric observations of Betelgeuse in the optical domain, H, and K band \citepads{2010A&A...515A..12C} and in the K band \citepads{2014A&A...572A..17M}. The visibilities were reproduced up to 120~arcsec$^{-1}$.\\

We used RHD simulations computed with the CO$^5$BOLD code (COnservative COde for the COmputation of COmpressible COnvection in a BOx of L Dimensions, $L = 2, 3$,  \citeads{2012JCoPh.231..919F}). This code solves the coupled equations of compressible hydrodynamics and non-local radiation transport. The main characteristics of the four simulations we used are summarized in Table \ref{Tab:Charac_Simu}, and they are also described in details in \citetads{2011A&A...535A..22C}. Rotation is not included in these models.

\begin{table*}[ht!]
        \caption{Characteristics of the four RHD simulations model used to analyze our VLTI/PIONIER data of Betelgeuse \citepads{2011A&A...535A..22C}.}
        \label{Tab:Charac_Simu}
        \begin{tabular}{lllllllll}
                \hline\hline
                \noalign{\smallskip}
                Model & Simulated & Grid & M$_\mathrm{pot}$ & L & T$_\mathrm{eff}$ & R$_\star$ & $\log g$ & Opacity\\
                 & relaxed & (grid & (M$_\odot$) & (L$_\odot$) & (K) & (R$_\odot$) & & approx. \\
                 & time (yr) & points) & & & & & & \\
                \hline
                \noalign{\smallskip}
                st35gm03n07 & 5.0 & $235^3$ & 12 & $91\,932 \pm 1400$ & $3487 \pm 12$ & $830.0 \pm 2.0$ & $-0.335 \pm 0.002$ & gray\\
                st35gm03n13 & 7.0 & $235^3$ & 12 & $89\,477 \pm 857$ & $3430 \pm 8$ & $846.0 \pm 1.1$ & $-0.354 \pm 0.001$ & non-gray\\
                st36gm00n04 & 6.4 & $255^3$ & 6 & $24\,211 \pm 369$ & $3663 \pm 14$ & $386.2 \pm 0.4$ & $0.023 \pm 0.001$ & gray\\
                st36gm00n15 & 3.0 & $401^3$ & 6 & $24\,233 \pm 535$ & $3710 \pm 20$ & $376.7 \pm 0.5$ & $0.047 \pm 0.001$ & gray\\
                \hline
        \end{tabular}
\end{table*}

We used several hundreds of snapshots of those simulations, each one representing the temporal evolving convection pattern on the stellar surface. Intensity maps are computed using the 3D pure-LTE transfer code Optim3D \citepads{2009A&A...506.1351C}. For each epoch, we computed intensity maps in the spectral channels of PIONIER at the desired spectral resolution. Following \citetads{2009A&A...506.1351C,2010A&A...515A..12C} and \citetads{2014A&A...572A..17M}, and since we do not know the orientation of the simulation relatively to the star on the plane of the sky, we rotated each intensity image around its center, over 18 positions between 0$^\circ$ and 180$^\circ$.
We then scaled the star to the apparent size, as derived from our LDD and Gaussian hotspot model (Tables \ref{Tab:Result_fit_hotspot} and \ref{Tab:fit_Nov14}). To obtain the interferometric observables, we computed the Fourier transform of these images and derived the visibility and the closure phase at the ($u, v$) points that we sampled with our observations.\\

\begin{figure*}
        \centering
        \resizebox{\hsize}{!}{\includegraphics{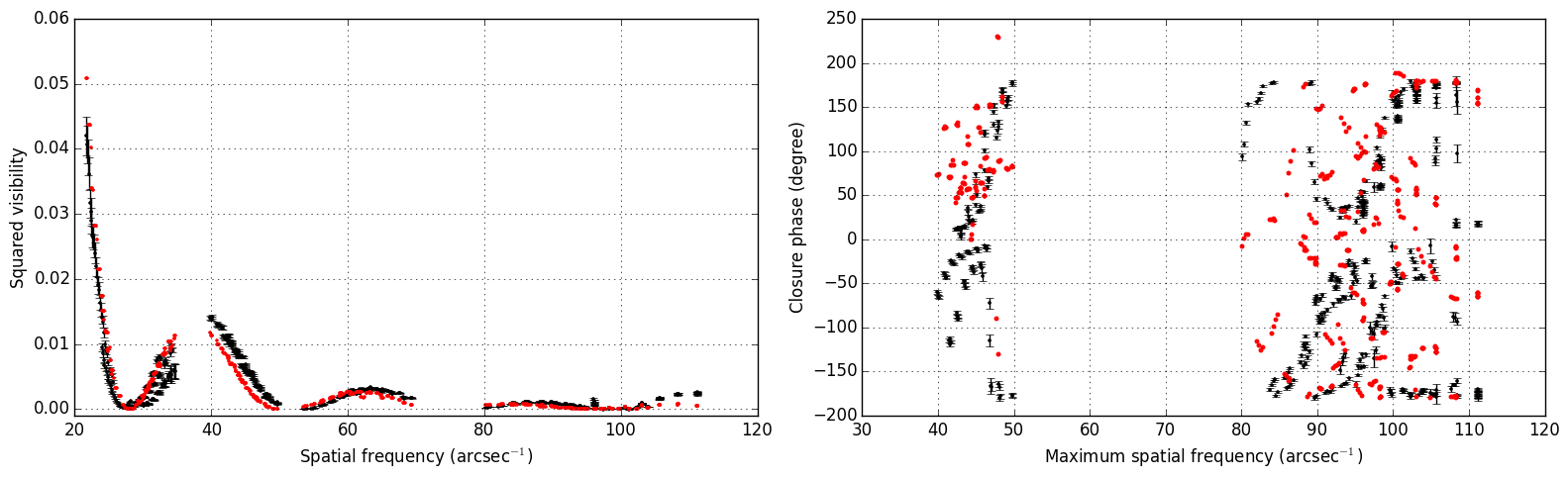}}\\
        \resizebox{\hsize}{!}{\includegraphics{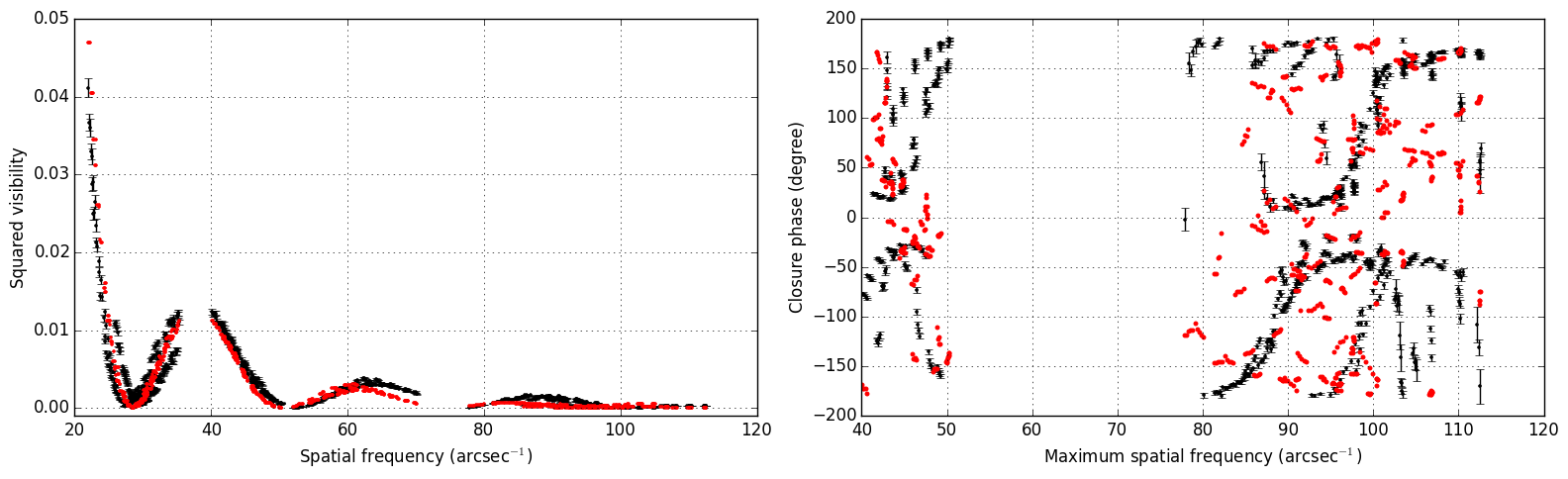}}\\
        \resizebox{\hsize}{!}{\includegraphics{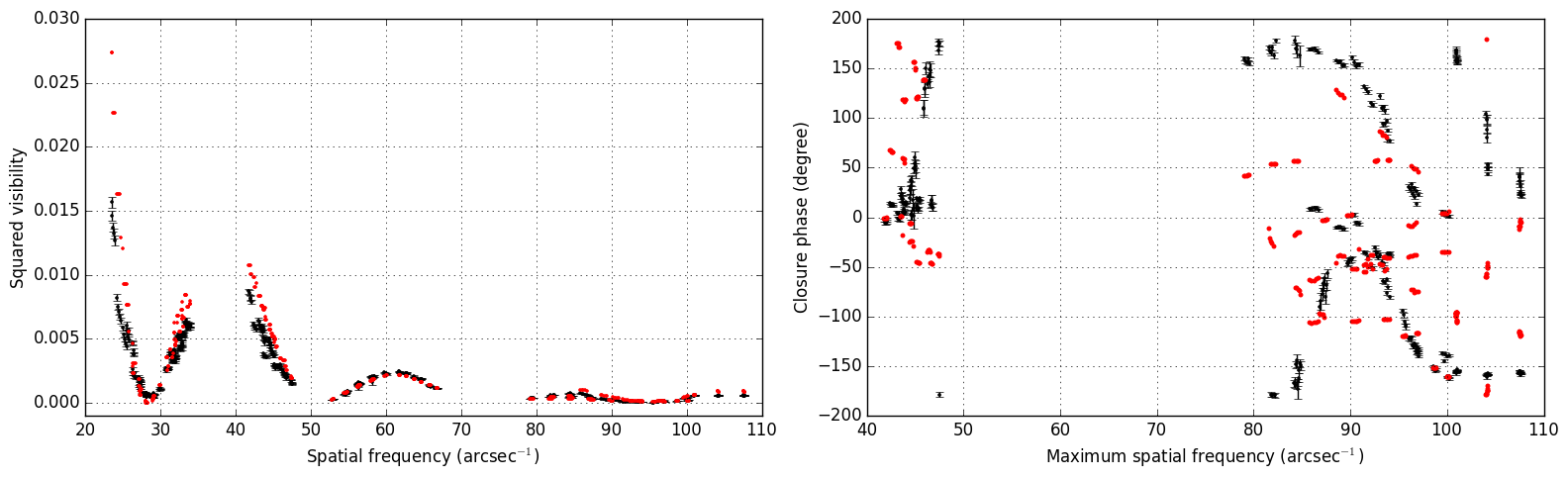}}\\
        \resizebox{\hsize}{!}{\includegraphics{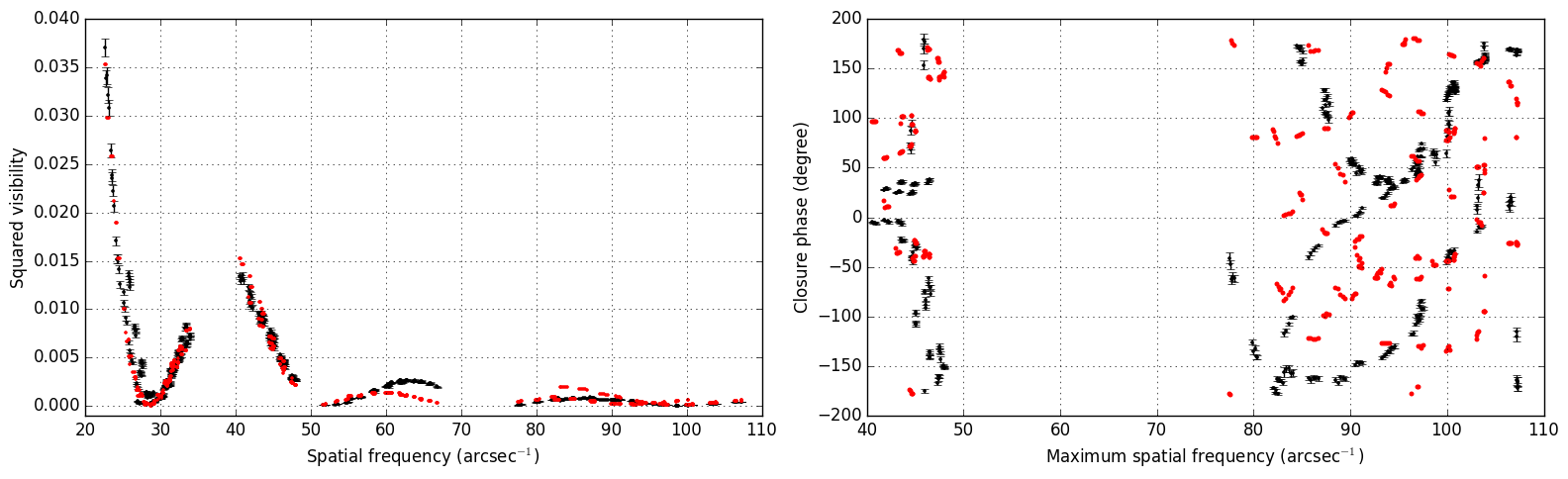}}\\
        \caption{Comparison of the squared visibilities (\textit{left}) and closure phases (\textit{right})  of the VLTI/PIONIER data in black and the best snapshot of the RHD simulation in red for each epoch. Only spatial frequencies greater than 51 arcsec$^{-1}$ were considered. Only the visibilities are fitted. The four rows corresponds to the 2012 January epoch (\textit{first row}), 2013 February epoch (\textit{second row}), 2014 January epoch (\textit{third row}), and 2014 November epoch (\textit{fourth row}). \label{Fig:v2_cp_simu}}
\end{figure*}

We compared these values to our visibilities by considering only spatial frequencies greater than 51~arcsec$^{-1}$, i.e., the domain not affected by the hotspot we had observed. The results of these fits are summarized in Table \ref{Tab:Result_Simu} and the observables are represented in Fig. \ref{Fig:v2_cp_simu}. While the visibilities are quite well reproduced, this is not the case for the closure phases. For this observable, even the shape is not similar.

\begin{table}[ht!]
        \caption{Best-fitted RHD simulations to the four epochs of VLTI/PIONIER data. Only spatial frequencies greater than 51 arcsec$^{-1}$ were considered. Only the visibilities are fitted.}
        \label{Tab:Result_Simu}
        \centering
        \begin{tabular}{lll}
                \hline\hline
                \noalign{\smallskip}
                Epoch & Name & $\tilde{\chi}^2$ \\
                \hline
                \noalign{\smallskip}
                2012-01 & dst36g00n04 & 906 \\
                2013-02 & dst35gm03n07 & 682 \\
                2014-01 & dst36g00n04 & 391 \\
                2014-11 & dst35gm03n07 & 679 \\
                \hline
        \end{tabular}
\end{table}

In the past, RHD simulations fitted  the interferometric observables of Betelgeuse well \citepads{2009A&A...506.1351C,2010A&A...515A..12C,2014A&A...572A..17M} but this is not the case for our four epochs of VLTI/PIONIER observations that  monitor the H-band photosphere. Considering that this is also the first time that a large persistent hotspot has been observed (Sect. \ref{SubSect_HotSpot}), we propose that this large feature is modifying the signature of the convective pattern at the surface of the star in a way the RHD simulation cannot reproduce. The regular convective pattern may still be present on the photosphere, but with a different activity. The interferometric signal is dominated by the large hotspot.

\section{Discussion \label{Sect_Discussion}}

\subsection{Photocenter displacement}

The presence of a non-centered hotspot on the photosphere of Betelgeuse is displacing its photocenter. We computed its position numerically from the intensity maps that were derived from our LDD and Gaussian hotspot model (Sect. \ref{SubSect_HotSpot}). The value are presented in Table \ref{Tab:photocenter}.

\begin{table}[ht!]
        \caption{Offset position and displacement of the photocenter of Betelgeuse, taking into consideration our LDD and Gaussian hotspot model (Sect. \ref{SubSect_HotSpot}).}
        \label{Tab:photocenter}
        \centering
        \begin{tabular}{l l l l}
                \hline\hline
                \noalign{\smallskip}
                Epoch & $\Delta$R.A. (mas) & $\Delta$Dec. (mas) & $\Delta$r (mas) \\
                \hline
                \noalign{\smallskip}
                2012-01 & -0.87 & -0.29 & 0.91 \\
                2013-02 & -0.43 & -0.13 & 0.45 \\
                2014-01 & -2.19 & 0.44 & 2.23 \\
                2014-11 & 1.22 & 0.47 & 1.31 \\
                \hline
        \end{tabular}
\end{table}

The value we derived corresponds to $\sim 10\%$ of the parallax of Betelgeuse \citepads{2008AJ....135.1430H}. Particularly between January and November 2014, the displacement reaches 3.41~mas. This value is more than six times larger than the photocenter displacement that was caused by related convective surface structures in the optical, where the contrast is up to five times  greater \citepads{2011A&A...528A.120C}. This is further evidence that there is an ingredient missing in RHD simulations.

Until now, such a  large hotspot has only been observed on Betelgeuse but the routine operation of the multi-baseline interferometer should allow more observations of multiple position angles of the first two lobes of the visibility function of nearby red supergiants.

\subsection{Convection and mass loss}

Giant convective cells were predicted on RSG by \citetads{1975ApJ...195..137S} and have been numerically reproduced in RHD simulations \citepads{2009A&A...506.1351C,2010A&A...515A..12C,2011A&A...535A..22C}. \citetads{2007A&A...469..671J} propose that such structures could be at the origin of mass loss by lowering the effective gravity of the star, in conjunction with radiative pressure on molecular lines. The reconstructed high-dynamic range image of \citetads{2009A&A...508..923H} in the H band from IOTA interferometric data already showed  the presence of two bright spots that were identified as the tops of convective cells by \citetads{2010A&A...515A..12C} using RHD simulation. However, RHD simulations cannot reproduce the observed VLTI/PIONIER low-frequency data (i.e., very large structures) with, in particular, very high and "fast" variable (about 1 year) surface contrast. This indicates that there is a physical  ingredient missing in the RHD simulations.

In our four epochs of observations, the contribution of the hotspot to the total flux equals respectively 7\%, 3\%, 19\%, and 6\%. These values are in agreement with previous observations on Betelgeuse and other stars of \citetads{2009A&A...508..923H,2010ASPC..425..140K} and predictions from \citetads{1975ApJ...195..137S}. This high level of brightness is a strong argument in favor of the hypothesis for the triggering the mass loss with convection by a combination of a lowering of the effective gravity, thanks to convection, and of radiative pressure on molecular lines \citepads{2007A&A...469..671J}.

It is particularly interesting to notice that in the intensity images (Fig. \ref{Fig:Int_maps}), part of the flux is coming from outside the stellar disk owing to the Gaussian nature and edge position of the best-fitted hotspot. It is this particular feature which is causing the characteristic shape of the visibility curve at low spatial frequencies. This can be interpreted in two different ways: we could be observing a convective cell at the limb of the star, making the star appear larger in this single direction (see, e.g. the  simulations of \citeads{2011A&A...535A..22C,2012JCoPh.231..919F} with smaller limb convective cells). This contribution outside of the disk could also be a plume that is emerging from the photosphere, which is thus  a stronger mass loss event. Both interpretations do not exclude each other since a limb-located convective cell could actually trigger a plume. In Paper III, VLT/SPHERE visible polarimetric observations with ZIMPOL (Zurich imaging polarimeter) revealed an arc located 3R$_\star$ northeast from the star. It was identified as an incomplete dust shell. As we do not know when the hotspot we detected appeared, we cannot infer if these two features are related but this hypothesis cannot be excluded. In addition, we must not forget that we do not have dynamic information and that, depending on the velocity field of the event, an important part of the material may fall back onto the photosphere. \citetads{2009A&A...503..183O,2011A&A...529A.163O} observed both upwards and downwards motions of CO in the MOLsphere of the star at 1.3 stellar radii. \\

Of course we cannot exclude other phenomena that might cause the star to appear elongated in one particular direction. Without contemporaneous imagery of the very close circumstellar environment near the infrared photosphere, it is impossible to completely disentangle alternate hypotheses. The recently installed instrument VLT/SPHERE \citepads{2008SPIE.7014E..41B} is already providing this kind of near infrared observations. However, for  almost the last 20 years of interferometric observations in the near infrared, using various instruments and ($u, \, v$) plane sampling, the Betelgeuse LDD diameter variation did not exceed 3~mas (Fig. \ref{Fig:Diam_Evol}). Moreover, our simple model of a Gaussian hotspot on an LDD reproduces both the visibilities and closure phase, at low spatial frequencies, the latter being mostly sensitive to asymmetries. Therefore, we consider that the departure from the disk structure for the near-infrared photosphere is unlikely.

\subsection{Counterpart in the AAVSO observations}

Betelgeuse has been regularly observed by the members of the American Association of Variable Star Observers (AAVSO). Using the \texttt{VSTAR}\footnote{Freely available online at \url{https://www.aavso.org/vstar-overview}} program, we downloaded and analyzed their visual observations between 1995 and 2014. With custom Python routines, we removed outlying points using the sigma-clipping technique. Then, we averaged the data over bins of ten days.
        
To estimate the period of the light curve, we made use of the implementation of the date-compensated discrete Fourier transform (DCDFT) algorithm in VSTAR. Our VLTI/PIONIER observations were obtained between January 2012 and November 2014. Therefore, we considered the AAVSO data that was obtained before, and after January 2012 separately. It is impossible to determine when the bright spot appeared, but \citetads{2009A&A...503..183O,2011A&A...529A.163O} and \citetads{2014A&A...572A..17M} did not observe this feature in their datasets that cover the January 2008 to January 2011 epoch. For the epoch before 2012, we derive a main period of $423.59 \pm 39$~d and of $422.01 \pm 88$~d after 2012. The presence of the large hotspot seems not to affect the main radial pulsation period. We could not see an influence on the long secondary period (LSP) of 2100 days \citepads{2010ApJ...725.1170S}, which is supposed to be caused by the convective cells' turnover time, as our observations cover less than 1 100 days.

\subsection{Comparison with TBL/Narval spectropolarimetric observations}

Betelgeuse was observed at the TBL (Telescope Bernard Lyot) between November 2013 and April 2015 with the spectropolarimeter Narval \citep{Auriere2016}. The linear polarization in spectral lines is ten times stronger than circular polarization, this latter being due to the Zeeman effect. 
It is clear that this linear polarization originates from the depolarization of the continuum, itself linearly polarized by Rayleigh scattering. Indeed, it is necessary to first polarize the continuum, and then to depolarize it through the absorption of a polarized photon and the re-emission of an unpolarized one. Any other source of continuum polarization that, as Mie scattering, may take place farther away, will not affect the observed line (de-)polarization.

Even if depolarization does take place, the azimuthal symmetry of the stellar disk may cancel it out in the observed profiles. 
The actual observation of a depolarization signal is interpreted as a breaking of this symmetry owing to brightness inhomogeneities in Betelgeuse: namely bright spots. The shape of the linearly polarized signal  thus depends on the position of such bright spots over the disk of Betelgeuse. The authors proposed an analytical model to map these spots over the disk (with an ambiguity of 180$^\circ$ in the position angle). From the shape (double Gaussian) of the linearly polarized signal, two spots are inferred. Depending on the different observation dates, the location of both spots is found to change. Indeed, before October 2014 the spots are located on the eastern limb for the brightest and on the southwestern limb for  the other, while in October 2014 the signal evolves with an apparent merging of the spots. To ensure continuity for the latter snapshots, the authors decided to choose the solution with the brightest spot on the southeastern limb and the other still on the southwestern limb. However, taking the intrinsic 180$^\circ$ ambiguity on the PA for each spot into consideration, the solution with a bright spot in the north-west quadrant is also valid and might match the evolution of the interferometric signal that we observed between January and November 2014 (Tables \ref{Tab:Result_fit_hotspot} and \ref{Tab:fit_Nov14}, and Fig. \ref{Fig:Int_maps}).

\section{Conclusion}

We observed the nearby RSG Betelgeuse with the four-telescope instrument VLTI/PIONIER during four epochs in  January 2012, February 2013 , January 2014,  and November 2014. In each dataset, by considering the first lobe of the visibility function, we observe that the star does not seem to have the same diameter, depending on the sampled direction on the plane of the sky. An LD elliptical model could not account for the closure phase signal. On the contrary, an LDD and Gaussian hotspot model converge on a solution that reproduces both observables and is consistent with spectropolarimetric observations with TBL/Narval \citep{Auriere2016}. 

We propose that this large structure may be part of the process that triggers the mass loss in RSG. Indeed, the best-fitted limb position for the Gaussian hotspot comes with light that is distributed outside  the stellar disk, and it is   that  allows the model to reproduce the interferometric observables. The large hotspot we used to interpret our observations is not predicted by RHD simulations. Moreover, the presence of this photospheric feature seems to affect the smaller convective cells by modifying their distribution and characteristic sizes. Currently, 3D convective simulations do not include rotation or magnetic effects. These phenomena, or another ingredient, could help to reproduce the VLTI/PIONIER observations, thus producing the missing atmospheric extension of these simulations. This type of event could be associated with an episodic mass loss as the basis of the clumpy environment observed in Papers I and II.\\

The photosphere of Betelgeuse and its CSE should now be monitored to look for evolution and consequences of this feature. We expect to find these types of structures  on other close RSG.

\begin{acknowledgements}
We are grateful to ESO's Director-General, Prof. Tim de Zeeuw, for the allocation of observing time to our program, as well as to the Paranal Observatory team for the successful execution of the observations.
This research received the support of PHASE, the high-angular resolution partnership between ONERA, Observatoire de Paris, CNRS, and University Denis Diderot Paris 7.
We acknowledge financial support from the Programme National de Physique Stellaire (PNPS) of CNRS/INSU, France.
Support for this work was provided by NASA through grant number HST-GO-12610.001-A from the Space Telescope Science Institute, which is operated by AURA, Inc., under NASA contract NAS 5-26555. 
We used the SIMBAD and VIZIER databases as well the Aladin sky atlas at the CDS, Strasbourg (France)\footnote{Available at \url{http://cdsweb.u-strasbg.fr/}}, and NASA's Astrophysics Data System Bibliographic Services.
This research has made use of Jean-Marie Mariotti Center's \texttt{Aspro}\footnote{Available at \url{http://www.jmmc.fr/aspro}} service, of the \texttt{LITpro}\footnote{Available at \url{http://www.jmmc.fr/litpro}} software (co-developped by CRAL, LAOG and FIZEAU) and of the \texttt{SearchCal} service\footnote{Available at \url{http://www.jmmc.fr/searchcal}} (co-developed by FIZEAU and LAOG/IPAG).
This research made use of IPython \citep{PER-GRA:2007} and Astropy\footnote{Available at \url{http://www.astropy.org/}}, a community-developed core Python package for Astronomy \citepads{2013A&A...558A..33A}.
We acknowledge with thanks the variable star observations from the AAVSO International Database that was contributed by observers worldwide and used in this research.
We would like to thank the anonymous referee whose suggestions and comments led to improvements in this article.
\end{acknowledgements}

\bibliographystyle{aa}
\bibliography{./biblio}

\listofobjects

\Online

\end{document}